\documentclass[a4paper]{article}
\usepackage[margin=1in]{geometry}
\usepackage[english]{babel}
\usepackage[utf8x]{inputenc}
\usepackage{amsmath}
\usepackage{graphicx}
\usepackage{float}
\usepackage[colorinlistoftodos]{todonotes}
\usepackage{authblk}
\usepackage{setspace}
\usepackage{multicol}
\usepackage{caption}
\usepackage{breqn}
\usepackage{lipsum}
\usepackage[utf8x]{inputenc}
\usepackage{tabularx}
\usepackage{nomencl}
\usepackage{subcaption}
\usepackage{tablefootnote}

\begin{document}
\begin{flushright}
{\it{Accepted for publication in Journal of Fluids and Structures, Dec 2017}}
\end{flushright}
\begin{large}

\begin{center}
{\LARGE{\bf{Role of Skin Friction Drag during Flow-Induced Reconfiguration of a Flexible Thin Plate}}}
\end{center}
\begin{small}
\begin{center}
Awan Bhati\(^\dagger\), Rajat Sawanni\(^\dagger\), Kaushik Kulkarni, Rajneesh Bhardwaj*\\
Department of Mechanical Engineering\\
Indian Institute of Technology Bombay \\
Mumbai 400076, India\\
\(^\dagger\)equal contributors\\
*Corresponding author (email: rajneesh.bhardwaj@iitb.ac.in)\\
Phone: +91 22 2576 7534, Fax: +91 22 2572 6875
\end{center}
\end{small}

\begin{abstract}
We investigate drag reduction due to the flow-induced reconfiguration of a flexible thin plate in presence of skin friction drag at low Reynolds Number. The plate is subjected to a uniform free stream and is tethered at one end. We extend existing models in the literature to account for the skin friction drag. The total drag on the plate with respect to a rigid upright plate decreases due to flow-induced reconfiguration and further reconfiguration increases the total drag due to increase in skin friction drag. A critical value of Cauchy number ($Ca$) exists at which the total drag on the plate with respect to a rigid upright plate is minimum at a given Reynolds number. The reconfigured shape of the plate for this condition is unique, beyond which the total drag increases on the plate even with reconfiguration. The ratio of the form drag coefficient for an upright rigid plate and skin drag coefficient for a horizontal rigid plate ($\lambda$) determines the critical Cauchy number ($Ca_{cr}$). We propose modification in the drag scaling with free stream velocity ($F_{x}$ ${\propto}$ $U^{n}$) in presence of the skin friction drag. The following expressions of $n$ are found for $0.01 \leq Re \leq 1$, $n = 4/5 + {\lambda}/5$ for 1 $\leq$  $Ca$ $<$ $Ca_{cr}$ and $n = 1 + {\lambda}/5$ for $Ca_{cr} \leq Ca \leq 300$, where $Re$ is Reynolds number. We briefly discuss the combined effect of the skin friction drag and buoyancy on the drag reduction. An assessment of the feasibility of experiments is presented in order to translate the present model to physical systems.

\end{abstract}
\section{Introduction}
The interaction between flexible, thin structures and fluid flow has several applications in biology and engineering systems. Some of the examples include, flow-induced reconfiguration of flexible plants or aquatic vegetation for drag reduction \cite{Vogel_1996, Langre}, flapping and bending of flexible sheets for applications in energy harvesting and sails \cite{Shelley1}, biolocomotion at microscale (e.g flagellum, cilia) \cite{Purcell} and bending of elastic fibers in microfluidics \cite{Shelley2,Wexler}.  

The classical form drag scaling  (\(F∝ U^2\)) of a flexible plant subjected to a free stream velocity was modified by Vogel \cite{Vogel_1996} to account for the effect of streamlining to obtain \(F∝ U^{2+V}\), where \(V\) is the Vogel exponent. In a series of papers, Alben et al. \cite{Alben1,Alben2} combined theoretical and experimental methods for investigating the drag reduction of a compliant fiber tethered at its midpoint. In these studies, authors introduced the ratio of fluid kinetic energy to elastic potential energy, as a dimensionless parameter to quantify the drag reduction at large Reynolds number. They concluded that the drag force scales as \(F∝ U^{4/3}\) ($V = -2/3$), consistent with the inclusion of the Vogel exponent in the classical drag scaling. In the context of aquatic vegetation, the range of Vogel exponent is $[-0.2, -1.2]$ in the reported measurements \cite{Langre2}. Zhu \cite{Zhu} used an immersed boundary method based two-dimensional viscous flow solver to model the experiments carried out by Alben et al. \cite{Alben1}. This numerical study showed \(U^{2}\) and \(U^{4/3}\) scaling at Reynolds number $Re = 10$ and $Re = 800$, respectively, where $Re$ is Reynolds number based on the length of the fiber. 

Gosselin et al. \cite{Gosselin} presented a combined experimental and theoretical study for the drag reduction of a thin plate tethered at its midpoint and circular disk with cuts along the radius tethered at its center. The variation of measured drag with respect to a rigid upright plate (reconfiguration number) with Cauchy number (ratio of hydrodynamic loading to the restoring force due to stiffness) collapsed on a single curve and agreed well with the predictions of a model based on Euler-Bernoulli beam theory. They showed that the streamlining is more important at the onset of reconfiguration while the area reduction dominates at larger deformation. Luhar and Nepf \cite{Luhar1} extended previous models \cite{Alben1, Gosselin} and considered a flexible plate tethered at one end in presence of buoyancy. Their measurements of the drag force with varying flow velocities compare well with the theoretical predictions. In addition, this study \cite{Luhar1} showed that the exponent $a$ of the drag scaling, \(F∝ U^a\), is lesser than 1 (\(a < 1\)) in presence of dominating buoyancy. 

Henriquez and Barrero-Gil \cite{Hen} investigated the effect of sheared incoming inflow on the reconfiguration of the flexible plate and showed that the Vogel exponent strongly depends on the profile of the sheared flow. Leclercq and de Langre {\cite{Leclercq}} developed a theoretical framework for finding the Vogel exponent of a flexible beam in the limit of large velocity flows. The model presented in this study {\cite{Leclercq}}  included non-uniformity in the flow as well as structural parameters. Authors concluded that the predicted Vogel exponent for large loading is around -1, which is consistent with previous experimental observations \cite{Langre2}. 

Most of the previously-mentioned theoretical models on the drag reduction of the flexible thin structures considered large Reynolds number at which the skin friction drag is negligible. Recently, two studies considered the effect of sheared flow in presence of a wall {\cite{Hen,Leclercq}}, however, these studies did not consider the effect of the skin friction drag. To the best of our knowledge, the effect of the skin friction drag on a tethered thin structure subjected to a free stream at low Reynolds number is not reported thus far. To this end, the objective of the present work is to extend the model of Gosselin et al. \cite{Gosselin} or Luhar and Nepf \cite{Luhar1} to investigate the effect of the skin friction drag on a thin plate for a wide range of Reynolds number, {\(0.01\leq Re \leq 10^8\)}.

The layout of this paper is as follows. The mathematical model is presented in section 2, and the drag reduction due to the flow-induced reconfiguration is discussed in absence and presence of the skin friction drag in sections 3.1 and 3.2, respectively. A scaling analysis of the drag force in presence of the skin friction drag and a feasibility study of experiments are presented in sections 3.3 and 3.4, respectively.

\section{Mathematical Model}
We extend the model described by Luhar and Nepf \cite{Luhar1} to account for the skin friction drag. The model presented by Luhar and Nepf \cite{Luhar1} was built upon the works of Alben et al. \cite{Alben1} and Gosselin et al. \cite{Gosselin}. We consider a thin, flexible, buoyant and inextensible plate in two-dimensional coordinates. The plate is tethered at one end and subjected to a uniform free stream (Fig. 1). The length, width, and thickness of the plate are \(l\), \(b\) and \(t\) respectively. The plate is considered very thin such that $t{{\ll}}l$ and $t{{\ll}}b$. The following subsections present the details of the modeling and all symbols used in the paper are defined in Table 1. 

\subsection{Governing equation and boundary conditions}

As shown in Fig. 1, we use a curvilinear axis attached to one end of the plate in which \(s\) is the distance from the base and \(\theta\) is the angle between the plate and vertical at any given point. The tip of the plate is located at \(s=l\). The form drag force per unit plate length at a curvilinear distance \(s\) on the plate is given by quadratic law \cite{Gosselin,Luhar1}, 
\begin{dmath}
f_{Dp} = (1/2)\rho C_{Dp} b U^{2}\textrm{cos}^{2}\theta
\end{dmath}
where \(\rho\) is the density of the fluid, \(C_{Dp}\) is the form drag coefficient for an upright rigid plate mounted perpendicular to the free stream. \(U\)cos\(\theta\) is the normal flow velocity to the plate at any point. 

Similarly, the skin friction drag force is represented using tangential velocity \(U\)sin\(\theta\) at any point on the plate and is given by, 
\begin{dmath}
f_{Ds} = (1/2)\rho C_{Ds} b U^{2}\textrm{sin}^{2}\theta
\end{dmath}
where \(C_{Ds}\) is the skin friction drag coefficient for a rigid horizontal plate mounted parallel to the free stream. \(U\)sin\(\theta\) is the flow velocity along the deformed beam at any point. Note that \(\theta = 0^o\) and \(\theta = 90^o\) at all points on the plate correspond to a rigid plate which is perpendicular and parallel to the direction of the free stream flow, respectively. The form drag force is the maximum in the former while the skin friction drag force is the maximum in the latter.

The buoyant force acts in the vertical direction at any point on the plate and its magnitude per unit plate length is expressed by \cite{Luhar1},  
\begin{dmath}
{f_B = (\rho-\rho_b)gbt}
\end{dmath}
where \( \rho_b \), \(b\) and \(t\) are plate density, width and thickness, respectively. We consider these forces for the region \(s\geq s^*\) shown in Fig. 1, where \(s^*\) is an arbitrary position along the plate. The hydrodynamic and buoyant forces are balanced by the restoring force acting at the curvilinear position \(s=s^*\), given by Euler-Bernoulli beam equation as follows, 
\begin{dmath}
V = -EI \frac{d^2\theta}{ds^2}
\end{dmath}
where, {{\it V}} is the restoring force normal to the plate due to stiffness, {{\it E}} is Young's Modulus and {{\it I}} is second moment of inertia, \(I=bt^3/12\). On balancing the forces along a normal plane to the plate, the following governing equation is obtained,
\begin{dmath}
V^*(s^*)+\int_{s^*}^{l} f_B \textrm{sin}(\theta^*(s^*)) ds = 
\int_{s^*}^{l} \textrm{cos}(\theta(s)-\theta^*(s^*))f_{Dp} (\theta(s)) ds+\int_{s^*}^{l} \textrm{sin}(\theta(s)-\theta^*(s^*))f_{Ds} (\theta(s)) ds
\end{dmath}
On substituting expressions of \(f_{Dp}\), \(f_{Ds}\) and \(f_{B}\) in eq. 5 and further simplification with non-dimensionalization \cite{Luhar1} result in the following governing equation: 
\begin{dmath}
{-\frac{d^2\theta}{d\hat{s}^2}|_{\hat{s}^{*}} +B(1 -\hat{s}^{*})\textrm{sin}(\theta^{*})} = \\{Ca\bigg( \int_{\hat{s}^*}^{1} \textrm{cos}(\theta(s)-\theta^*(s^*)) \textrm{cos}^2(\theta) d\hat{s} }+{\lambda\int_{\hat{s}^*}^{1} \textrm{sin}(\theta(s)-\theta^*(s^*)) \textrm{sin}^2(\theta) d\hat{s}\bigg)}
\end{dmath}
The first, second, third and fourth term in the above equation represent rigidity, buoyancy, form drag and skin friction drag, respectively. Note that eq. 6 without skin friction drag was reported by Luhar and Nepf \cite{Luhar1}. The non-dimensional parameter {\it{B}} represents the ratio of restoring force due to buoyancy and the restoring force produced by stiffness \cite{Luhar1} and is given by,
\begin{dmath}
{B=\dfrac{(\rho-\rho_b) g btl^3}{EI}}
\end{dmath}
Cauchy number, $Ca$, represents the ratio of hydrodynamic force experienced to the restoring force produced by stiffness \cite{Luhar1} and is expressed as follows,
\begin{dmath}
{Ca={\dfrac{1}{2}}\dfrac{\rho C_{Dp} b l^3 U^2}{EI}},
\end{dmath}
and {$\lambda$} is the ratio of the skin friction drag coefficient for a rigid horizontal plate mounted parallel to the free stream to the form drag coefficient for an upright rigid plate mounted perpendicular to the free stream, defined as
\begin{dmath}
{\lambda = \dfrac{C_{Ds}}{C_{Dp}}}
\end{dmath}
The constant $\lambda$ accounts for the effect of Reynolds number ($Re$) in the model. $Re$ is based on the free stream velocity and length of the plate and is defined as follows,
\begin{dmath}
Re = \dfrac{\rho U l}{\mu}
\end{dmath}
where $\mu$ is dynamic viscosity. Note that \(Re\) and \(Ca\)  are not entirely independent variables. Using eq. 8, eq. 10 and definition of $I$, the dependence of $Ca$ on $Re$ is expressed as follows,
\begin{dmath}
Ca= 6{\dfrac{C_{Dp}Re^2\mu^2}{\rho E}}\dfrac{a_{r}^3}{l^2},
\end{dmath}
where $a_{r}$ is aspect ratio of the plate ($a_{r} = l/t$). Eq. 11 suggests that an increase in $Re$ coincides with a larger increase in $Ca$ for a given fluid, plate material and plate geometry. The following boundary conditions are prescribed \cite{Alben1,Gosselin,Luhar1}. The base of the plate is considered as fixed on one end, \(\theta=0\) at \(s=0\). The tip of the plate is free i.e. \({d\theta}{/}{ds}\)\(=0\) at $s=l$.

\subsection{Quantities for quantifying drag reduction}

In order to quantify the drag reduction due to the reconfiguration, Gosselin et al. \cite{Gosselin} defined reconfiguration number, which is the ratio of drag in a configured state to drag on a rigid plate mounted perpendicular to the flow direction. This variable was called as effective length, $l_{e,p}{/}l$, by Luhar and Nepf \cite{Luhar1} and is expressed as,
\begin{dmath}
{\dfrac{l_{e,p}}{l} = \dfrac{\int_{0}^{l}(1/2)\rho  b U^2 C_{Dp} \textrm{cos}^3 \theta ds}{(1/2)\rho b U^2 l C_{Dp}}} =\int_{0}^{1}cos^3 \theta d\hat{s}
\end{dmath}
where $\hat{s} = s/l$ is non-dimensional coordinate along the plate length. The posture of the plate $\theta(s)$ is obtained by numerically solving eq. 6 and left hand side of eq. 10 is obtained using the calculated $\theta(s)$. Extending the definition of the effective length for the form drag \cite{Luhar1}, we similarly define effective length for the skin friction drag with respect to the upright rigid plate, $l_{e,s}{/}l$, as follows,
\begin{dmath}
{\dfrac{l_{e,s}}{l} = \dfrac{\int_{0}^{l}(1/2)\rho  b U^2 C_{Ds} \textrm{sin}^3 \theta ds}{(1/2)\rho b U^2 l C_{Dp}}} =\lambda\int_{0}^{1} sin^3 \theta d\hat{s}
\end{dmath}
The overall effective length with respect to the vertical rigid plate is the sum of the contributions of the form drag and skin friction drag, given by eqs. 10 and 11, respectively, and is expressed as, 
\begin{dmath}
{\dfrac{l_{e}}{l} = { \dfrac{l_{e,p}}{l} + \dfrac{l_{e,s}}{l}} =
\int_{0}^{1}cos^3(\theta)d\hat{s}}  +  \int_{0}^{1}\lambda \textrm{sin}^3(\theta) d\hat{s} 
\end{dmath}
Therefore, ${l_{e}}/{l}$ represents total drag on the plate with respect to that on an upright rigid plate (mentioned as {\it{total relative drag}} hereafter). The percentage of form drag \(\eta_p\) and skin friction drag \(\eta_s\) in the  total relative drag  are defined respectively as follows: 
\begin{dmath}
{\eta_p = \dfrac{l_{e,p}/l}{l_{e}/l}\times100};{\eta_s = \dfrac{l_{e,s}/l}{l_{e}/l}\times100}
\end{dmath}
The drag reduction due to the flow-induced reconfiguration is caused by area reduction and streamlining \cite{Langre,Gosselin,Luhar1}. The former is given by \cite{Luhar1},
\begin{dmath}
\dfrac{h}{l}= \int_{0}^{1} \textrm{cos}\theta d\hat{s}
\end{dmath}
The total effective length (eq. 12) is comprised of the area reduction (eq. 14) as well as  streamlining. Therefore, the streamlining can be quantified by the following equation, 
\begin{dmath}
\dfrac{l_{e,str}}{l} = \dfrac{l_{e}}{l} - \dfrac{h}{l}
\end{dmath}

\subsection{Numerical Methodology}

Eq. (6) is discretized using a second-order central difference scheme and is solved by an iterative shooting method. Results of grid size convergence study are plotted in Fig. 2(a) in which the plate is discretized with $N$ = 20, 50, 100 and 200 grid points and the effective length ($l_{e}{/}l$) is plotted as a function of Cauchy number ($Ca$) for these cases. The following parameters are used in these simulations, \(Re = 10^8\), \(C_{Dp} = 1.95\), \(\lambda = 0\) and \(B = 0\), where $Re$, $\lambda$ and $B$ are Reynolds number, ratio of the drag coefficients (eq. 9) and Buoyancy number, respectively. The values of $L_2$ error-norm for $N$ = 20, 50 and 100 with respect to $N$ = 200 are 0.18, 0.06 and 0.02, respectively. Therefore, $N$ = 100 grid points are used for the simulations presented in section 3.

In order to validate the current methodology, we compare present results with those of Luhar and Nepf \cite{Luhar1} using same parameters as those used for the grid size convergence study. The comparison presented in Fig. 2(b) is excellent and validates our model.

\subsection{Simulation Setup}
The simulations are carried out for a wide range of Reynolds number ({\it{Re}}) as well as Cauchy number ({\it{Ca}}). The range of {\it{Re}} and {\it{Ca}} considered in present work are [${10^{-2}}$, ${10^8}$] and [${10^{-3}}$, ${10^5}$], respectively. The different $Re$ cases considered in the present study are listed in Table 2. The value of $\lambda$, ratio of drag coefficients (\(C_{Ds}\) and \(C_{Dp}\)), at given {\it{Re}} is needed in the model. We estimate these drag coefficients using the expressions and data reported in the literature, described as follows. The skin friction drag coefficient, \(C_{Ds}\), is calculated using the following expressions reported in the literature \cite{Vogel_1996, Tom, White}.
\begin{dmath}
{C_{Ds}} = \left\{
        \begin{array}{ll}
 		\dfrac{4{\pi}}{ReS}[1-\dfrac{1}{S}({S^{2}}-S-\dfrac{5}{12})\dfrac{Re^2}{128}], & \quad Re \leq 1, \\
		\\
 		\dfrac{1.328}{\sqrt{Re}} + \dfrac{2.3}{Re}, & \quad 1 < Re \leq 100,\\
        \\
        \dfrac{1.328}{\sqrt{Re}}, & \quad Re > 100,
        \end{array}
    	\right.
\end{dmath}
where $S$ = 3.1954 -log($Re$) and \(C_{Ds}\) for $Re$ $\leq$ 1 in eq. 16 was obtained using Oseen's equations in Ref. \cite{Tom}. Eq. 18 includes a correction factor (2.3/$Re$) given in a perturbation theory \cite{Imai,White} for 1 $<$ $Re$ $\leq$ 100. The expression for $Re$ $\geq$ 100 in eq. 18 is Blasius boundary layer solution for a flat plate. Fig. 3(a) plots the variation of $C_{Ds}$ with $Re$ obtained using eq. 18 and the different cases of $Re$ considered are shown as symbols in this figure.

The expression of the form drag coefficient \(C_{Dp}\) for $Re$ $\leq$ 1 is calculated using the following expression \cite{Tom},
\begin{equation}
C_{Dp} = \dfrac{4{\pi}}{ReS^{*}}[1-\dfrac{1}{S^{*}}({{S^{*}}^{2}}+S^{*}+\dfrac{1}{4})\dfrac{Re^2}{128}],  \quad Re \leq 1,
\end{equation}
where $S^{*}$ = 2.1954 -log($Re$). The values of \(C_{Dp}\) for different $Re$ cases falling within range of $Re$ = [10, $10^8$] are taken from Refs. \cite{Luhar1, Vogel_1996} and are listed in Table 2. The variation of $C_{Dp}$ with $Re$ obtained using eq. 19 and reported by Vogel \cite{Vogel_1996} is plotted in Fig. 3(b).

\section{Results and Discussion}
Results obtained by the model presented in section 2 are organized in this section as follows. First, the drag reduction in absence of skin friction drag is quantified in section 3.1. Second, the effect of the skin friction drag on the drag reduction is quantified and is discussed in section 3.2.  Third, a scaling analysis of the effective length and drag force is presented in section 3.3. Finally, we present feasibility of experiments in section 3.4 which may help to translate the present model to physical systems. The limitations of the present model are discussed in section 3.5. 

\subsection{Drag reduction in absence of skin friction drag}

The drag reduction due to the flow-induced configuration in absence of the skin friction drag is explained in the literature \cite{Alben1, Alben2, Gosselin, Luhar1} by two factors, namely, area reduction and streamlining. In the present paper, we quantify these two factors, expressing them as \(h/l\) (eq. 16) and \(l_{e,str}/l\) (eq. 17), respectively, and the effective length (\(l_{e}/l\)) is the sum of these two factors. We consider a plate with finite stiffness in absence of skin friction drag at large Reynolds number, $Re = 10^8$ ($\lambda = 0$). Fig. 4 plots the variation of \(h/l\), \(l_{e,str}/l\) and \(l_{e}/l\) with respect to Cauchy number, \(Ca\), for this case. For 0.001 $<$ \(Ca\) $<$ 1, the rigidity of the plate dominates the fluid loading and thereby the plate does not reconfigure along the flow direction. Therefore, there is a negligible reduction in the total relative drag. The leftmost inset on the top of Fig. 4 plots the shape of the plate for \(Ca\) = 0.06. For 1 $<$ \(Ca\) $<$ 26, \(h/l\), \(l_{e,str}/l\) and \(l_{e}/l\) decrease with  \(Ca\) and the flow-induced reconfiguration reduces drag on the plate due to the area reduction as well as streamlining. The plate gets streamlined along the flow due to the reconfiguration and the middle inset on the top of Fig. 4 plots the shape of the plate for \(Ca\) = 26. 

For 26 $<$ \(Ca\) $<$ \(10^4\), \(h/l\) decreases with \(Ca\), showing further area reduction with \(Ca\). On the other hand, \(l_{e,str}/l\) starts increasing with \(Ca\), demonstrating that the effect of the streamlining is decreasing to reducing the drag, as compared to that in case of 1 $<$ \(Ca\) $<$ 26. However, the magnitudes of \(h/l\) and \(l_{e,str}/l\) at a given \(Ca\) are comparable, implying that the streamlining is equally important as area reduction in this range of \(Ca\). The shape of the plate at \(Ca = 920\) is plotted in the rightmost inset in Fig. 4. Overall, the total relative drag (\(l_{e}/l\)) for 1 $<$ \(Ca\) $<$ \(10^4\) decays with a power function of \(Ca\) ($l_{e}/l$ $\propto$ $Ca^{-1/3}$ \cite{Luhar1}).  

\subsection{Effect of Skin Friction Drag}

\subsubsection{Limiting Cases}
In a limiting case of a plate with very large stiffness (\(E\) $\rightarrow$ $\infty$, \(Ca\) $\rightarrow$ $0$, where $E$ is Young's Modulus), the skin friction drag is not important since \(\theta =\theta^{*}=0^{\circ}\) for  $s > 0$ on the plate (Fig. 1)  and last term in eq. 5 representing the skin friction drag is zero in this case. This case corresponds to an upright rigid plate to the direction of the flow. On the other hand, in the limiting case of a plate with zero stiffness (\(E\) $\rightarrow$ $0$, \(Ca\) $\rightarrow$ $\infty$), form drag is not important since \(\theta = \theta^* = 90^{\circ}\) for  $s > 0$ and third term in eq. 5 representing the form drag is zero in this case. This case corresponds to a horizontal plate aligned in the direction of flow.

\subsubsection{Baseline case at Reynolds Number, \(Re\) = 1}

In order to investigate the effect of the skin friction drag on the drag reduction, we consider a plate with finite stiffness at low Reynolds number, \(Re\) = 1. Fig. 5(a) plots the variation of the effective lengths, \(l_{e,p}/l\), \(l_{e,s}/l\) and  \(l_{e}/l\), that are given by eqs. 12, 13, 14, respectively, with respect to Cauchy number, \(Ca\). As discussed in section 2.2, \(l_{e,p}/l\) and \(l_{e,s}/l\) are the contribution of the form drag and skin friction drag in the  total relative drag, respectively, experienced by a flexible plate. The variation of \(l_{e}/l\) in Fig. 5(a) shows that the  total relative drag reduces for 1 $<$ \(Ca\) $<$ 27. This is due to the flow-induced reconfiguration, as explained in section 3.1. However, \(l_{e}/l\) further increases for 27 $<$ \(Ca\) $<$ \(10^3\), showing that the skin friction drag (\(l_{e,s}/l\)) dominates over the form drag \(l_{e,p}/l\) and the total relative drag starts increasing in this range of \(Ca\). \(l_{e}/l\) exhibits a minima at a critical Cauchy number (\(Ca_{cr}\) = 27), and both \(l_{e,s}/l\) and  \(l_{e,p}/l\) contribute to \(l_{e}/l\) at \(Ca_{cr}\). The shape of the plate at \(Ca_{cr} = 27\) is shown in the middle inset on top of Fig. 5(a). Therefore, there exists a critical Cauchy number $Ca_{cr}$ at which the total relative drag is minimum at a given Reynolds number and the configuration of the plate is unique. At larger Cauchy number, \(Ca\) $>$ \(Ca_{cr}\), the plate undergoes flow-induced deformation, however, with an increase in the total relative drag due to dominating skin friction drag. The corresponding configuration of the plate at \(Ca\) = 920 is nearly horizontal and is shown in the rightmost inset in Fig. 5(a). 

In order to quantify the area reduction and streamlining for \(Re\) = 1, we plot the variation of \(h/l\), \(l_{e,str}/l\) and \(l_{e}/l\) with respect to Cauchy number, \(Ca\), in Fig. 5(b). At critical Cauchy number (\(Ca = Ca_{cr}\)), the streamlining is negligible, $l_{e,str}/l \approx 0$ and \(l_{e}/l \approx h/l\) in Fig. 5(b), that also follows from eq. 17. Therefore, the projected shape of an upright rigid plate at critical Cauchy number represents the  total relative drag. For \(Ca > Ca_{cr}\), \(h/l\) and \(l_{e,str}/l\) decreases and increases with $Ca$, respectively, and \(l_{e}/l\) increases i.e. the total relative drag increases despite the area reduction due to the flow-induced reconfiguration. This implies that although the streamlining at \(Ca > Ca_{cr}\) reduces the form drag yet it increases the skin friction drag. Consequently, the total relative drag increases due to the net effect of the streamlining at \(Ca > Ca_{cr}\).

\subsubsection{Effect of Reynolds Number}

In order to further understand the effect of Reynolds number ($Re$) on the drag reduction, simulations are performed with different cases of $Re$ in the following range, 0.01 ${\leq}$ \(Re\) ${\leq}$ \(10^8\).  Fig. 6 compares the variation of effective length \(l_{e}/l\) with \(Ca\), obtained for several cases of \(Re\). We note that the non-monotonic variation of \(l_{e}/l\) with \(Ca\)  at \(Re\)  = 0.1 and 0.01 is similar to previously explained trend for \(Re=1\) in section 3.2.2. On the other hand, at larger Reynolds number, \(Re\) = 10, 100 and 1000, the total relative drag reduction behaves similar to that explained for \(Re\) = $10^8$ in section 3.1. Thus, existence of the total relative minimal drag is function of \(Re\) (or specifically $\lambda$, eq. 9) and \(Ca\). 

The values of the critical Cauchy number (\(Ca_{cr}\)) corresponding to the total relative minimum drag for \(Re\) = 0.01, 0.1 and 1.0 are 15.6, 17.8 and 27.2, respectively. Therefore, \(Ca_{cr}\) increases with increase in \(Re\), as plotted in Fig 7(a) and the shapes of the plate at \(Ca = Ca_{cr}\) and \(Re\) = 0.01, 0.1 and 1.0 are plotted in Fig 7(b). The plate with larger $Re$ shows larger bending due to larger \(Ca_{cr}\) (or lesser stiffness of the plate). Besides showing the relative minimum drag, these configurations also correspond to negligible streamlining, as discussed in section 3.2.2. 

For $Re$ $\rightarrow$ $10^8$ or $\lambda$ $\rightarrow$ $0$, a nearly horizontal plate (i.e. $\theta(s)$ $\rightarrow$ $90^o$) exhibits the total relative minimum drag, as shown in rightmost inset of Fig. 5(a). This also follows from eq. 14 and for this case, $Ca_{cr}$ $\rightarrow$ $\infty$. The effective length in the limiting case of a plate with zero stiffness  (\(E\) $\rightarrow$ $0$, \(Ca\) $\rightarrow$ $\infty$) asymptotes to $\lambda$ i.e. \(l_{e}/l\) $\rightarrow$ $\lambda$, as followed from eq. 14 (since $\theta(s) \rightarrow 90^\circ$). The computed values of \(l_{e}/l\) at \(Ca\) $\rightarrow$ $\infty$ correspond to $\lambda$ for all cases of $Re$ in Fig. 6. 

The percentages of form drag \(\eta_p\) and skin friction drag \(\eta_s\) in the total relative drag (eq. 15) are plotted with respect to \(Ca\) for different cases of \(Re\) in Fig. 8(a) and 8(b), respectively. Fig. 8 shows that the contribution of the form drag and skin friction drag increases and decreases with $Re$, respectively, at a given $Ca$. As expected, \(\eta_p\) and \(\eta_s\) decreases and increases with $Ca$, respectively, at a given $Re$. Fig. 8(b) shows that \(\eta_s\) is larger than 80{\%} for $Re$ ${\leq}$ $10$ for a wide range of $Ca$, confirming the importance of the skin friction drag at low Reynolds number during reconfiguration.

\subsubsection{Effect of Buoyancy}

Fig. 9 shows combined effect of the skin friction drag and buoyancy, and plots \(l_{e}/l\) with \(Ca\) for four cases described as follows. First, we introduce buoyancy at low \(Re\) and plot \(l_{e}/l\) for two cases, \(Re = 1\), \(B = 0\) and \(Re = 1\), \(B = 50\). Second, we introduce buoyancy at larger \(Re\) and plot \(l_{e}/l\) for the next two cases, \(Re = 10^8\), \(B = 0\) and \(Re = 10^8\), \(B = 50\). Comparing \(l_{e}/l\) for \(Re = 10^8\) shows that a larger \(Ca\) is required in presence of buoyancy for achieving the same amount of drag reduction. This is due to the fact that buoyancy produces a counteracting moment on the plate. Similar characteristics of \(l_{e}/l\) are noted at \(Re = 1\) in Fig. 9. The critical Cauchy number (\(Ca_{cr}\)) corresponding to the total relative minimum drag in presence of the skin friction drag is larger in presence of the buoyancy. 

\subsection{Scaling analysis}
It is well-established in the previous studies \cite{Alben1,Gosselin,Luhar1} that the drag force, $F_x$, due to the reconfiguration in the absence of skin friction drag scales as, \(F_{x} ∝ U^{4/3}\), where $U$ is the free stream velocity and the effective length scales as, $l_{e}/l\ \propto Ca^{-1/3} \propto U^{-2/3}$ \cite{Luhar1}. As plotted in Fig. 6, variation of \(l_{e}/l\) with $Ca$ in the presence of the skin friction drag at 0.01 ${\leq}$ \(Re\) ${\leq}$ ${1}$ is non-monotonic and therefore, we propose modified scaling laws for this range of $Re$. We present scaling for limiting cases and a plate with finite stiffness in the following subsections.

\subsubsection{Limiting Cases}

As discussed in section 3.2.1, the limiting case of a plate with very large stiffness (\(E\) $\rightarrow$ $\infty$, \(Ca\) $\rightarrow$ $0$) corresponds to a upright and rigid plate, and therefore, the drag force scales as $F_{x} ∝ U^{2}$. On the other hand, in the limiting case of a plate with zero stiffness (\(E\) $\rightarrow$ $0$, \(Ca\) $\rightarrow$ $\infty$), the plate is horizontal and aligned in the direction of the flow. In this case, drag force $F_x$ scales as (eq. 2),
\begin{dmath}
F_{x}\quad{\propto}\quad C_{Ds}U^2
\end{dmath}
Our numerical results verifies the scaling obtained in eq. 20, described as follows. Since \(l_{e}/l\) is the ratio of the drag force in the deformed configuration ($F_x$) to that for an upright rigid plate ($\rho C_{Dp}bU^2/2$, eq. 1), we obtain scaling of $F_x$ in terms of \(l_{e}/l\) and $U$ as follows,
\begin{dmath}
F_{x} \quad {\propto}\quad \dfrac{l_e}{l}C_{Dp}U^{2} 
\end{dmath}
Since the effective length \(l_{e}/l\) for the limiting case of zero stiffness (\(E\) $\rightarrow$ $0$, \(Ca\) $\rightarrow$ $\infty$) scales as $\lambda$ (section 3.2.3) and $\lambda$ = $C_{Ds}/C_{Dp}$ (eq. 9), eq. 21 is simplified as follows,
\begin{dmath}
F_{x} \quad {\propto}\quad {\lambda}C_{Dp}U^{2} \quad {\propto}\quad C_{Ds}U^2 
\end{dmath}
Thus, the scaling obtained in eq. 22 is consistent with eq. 20. Since $C_{Dp} ∝ {U}^{-0.85}$ for $Re \leq 1$ (Fig. 3), the drag force scales as, $F_{x} ∝ U^{1.15}$ at \(Ca\) $\rightarrow$ $\infty$ and 0.01 ${\leq}$ $Re \leq 1$.

\subsubsection{Plate of Finite Stiffness}

As discussed earlier, the effective length \(l_{e}/l\) decreases and increases with Cauchy number \(Ca\) for \(Ca\) $<$ \(Ca_{cr}\) and \(Ca\) $>$ \(Ca_{cr}\), respectively. In this section, we present scaling of \(l_{e}/l\) with \(Ca\) for these two regimes of \(Ca\) for 0.01 ${\leq}$ $Re \leq 1$. In order to obtain exponent $m$ for $l_{e}/l\ \propto Ca^{m}$, we fit linear profiles by least-squares method on log-log plot of \(l_{e}/l\) and \(Ca\) and thereby obtaining the exponent $m$ as slope of the linear fit. The profiles are fitted for 1 $\leq$ \(Ca\) $\leq$ \(Ca_{cr}\) and \(Ca_{cr}\) $\leq$ \(Ca\) $\leq$ 300, for the two regimes, respectively, as shown in Fig. 10 for $Re$ = 0.01, 0.1 and 1. All fits are obtained with $R^2$ value of larger than 0.98 and values of $m$ are shown for the fits in Fig. 10. 

In the first regime of \(Ca\) i.e. \(Ca\) $<$ \(Ca_{cr}\), $m$ is obtained as, $m$ = -0.13, -0.14 and -0.16 for $Re$ = 0.01, 0.1 and 1, respectively. $m$ is further fitted as a linear function of $\lambda$ as follows, $m = -0.283 + 0.175{\lambda}$, with $R^2$ = 0.99. Therefore, the scaling of \(l_{e}/l\) is expressed as follows,
\begin{dmath}
\dfrac{l_{e}}{l}\quad  {\propto}\quad Ca^{-0.283 + 0.175{\lambda}} 
\end{dmath}
We obtain the following scaling of $F_x$ with $U$ after substituting scaling of \(l_{e}/l\) with $Ca$ (eq. 23), $C_{Dp}$ with $Re$ for 0.01 ${\leq}$ $Re \leq 1$ ($C_{Dp} ∝ {U}^{-0.85}$, Fig. 3(b)) and $Ca$ with $U$ ($Ca ∝ C_{Dp}U^2$, eq. 8) in eq. 21. 
\begin{dmath}
F_{x} \quad {\propto} \quad U ^ {0.825 + 0.201{\lambda}} \quad {\approx} \quad U ^ {4/5 + {\lambda}/5}
\end{dmath}
Similarly, in the second regime \(Ca\) $>$ \(Ca_{cr}\), $m$ is obtained as, $m = -0.083 + 0.175{\lambda}$. In this regime, the following scaling of $F_x$ with $U$ is obtained,
\begin{dmath}
F_{x} \quad {\propto}\quad {U^{1.055 + 0.201{\lambda}}} \quad {\approx} \quad U ^ {1 + {\lambda}/5}
\end{dmath}

\subsubsection{Comparison among drag force scaling in different cases}

A comparison of exponent $n$ obtained for drag force scaling, $F_{x}$ $\propto$ $U^n$, among different cases, is presented in Fig. 11 using log-log plot of $F_{x} = kU^n$, where $k$ is a constant, same in all cases. Fig. 11 compares only $n$ for the different cases and therefore magnitude (or units) of $F_{x}$ and $U$ are not mentioned in the figure. The insets show the corresponding plate configuration and scaling of the cases. The flow-induced reconfiguration without the skin friction drag at $Re \geq 100$ reduces $n$ to 1.33 as compared to classical drag scaling for an upright rigid plate ($n=2$). The values of $n$ for a horizontal plate obtained using Oseen equations (0.01 ${\leq}$ $Re \leq 1$) and Blasius theory ($Re \geq 100$) are 1.15 and 1.5, respectively and are compared in Fig. 11. 

In presence of the skin friction drag for the flexible plate at 0.01 $\leq$ $Re$ $\leq$ 1, we approximate value of $\lambda$ as $\lambda$ $\approx$ 1 using data in Table 2 and we obtain $n = 1$ and $n = 1.2$ for \(Ca\) $<$ \(Ca_{cr}\) and \(Ca\) $\geq$ \(Ca_{cr}\), respectively. Therefore, the reconfiguration in presence of the skin friction drag results in the slowest increase ($n \approx 1$) in the drag with velocity for \(Ca\) $<$ \(Ca_{cr}\), as plotted in Fig. 11. Beyond the critical value of Cauchy number, the drag increases faster ($n = 1.2$) and the increase is roughly same as for a horizontal plate ($n = 1.15$) at 0.01 $\leq$ $Re$ $\leq$ 1. 

\subsection{Feasibility of Experiments}
We assess feasibility of experiments which may help to verify the present model. We examine fluids of different dynamic viscosities ($\mu$) - water ($\rho = 1000$ kg m$^{-3}$, $\mu = 1.0 \times 10^{-3}$ Pa s) and glycerol ($\rho = 1260$ kg m$^{-3}$, $\mu = 1.4$ Pa s) - and plates of different Young's Moduli ($E$) - silicone foam ($E = 5 \times 10^{5}$ Pa), polyurethane  ($E = 2.7 \times 10^{7}$ Pa), HDPE ($E = 9.3 \times 10^{8}$ Pa) and steel ($E = 2.2 \times 10^{11}$ Pa). In this context, Luhar and Nepf  \cite{Luhar1}  reported experiments for the flow-induced reconfiguration of thin plates of silicone foam and  HDPE in water at large $Re$.  We estimate length of the plate at the lowest total relative drag at $Re = 1$, which occurs at $Ca = Ca_{cr} = 27$, as discussed in section 3.2.2. Since the present model is valid for the thin plates, we consider $a_{r} = 100$ and $a_{r} = 1000$ for different combinations of the fluids and plate materials mentioned earlier. The calculated plate lengths using eq. 11 at critical $Ca$ are plotted in Fig. 12 and the largest plate length obtained is 2 m for glycerol with aspect ratio, $a_{r} = 1000$ i.e. thickness of 2 mm. Fig. 12 shows that the plate length increases with aspect ratio as well as with dynamic viscosity. As mentioned earlier, the plate length in Fig. 12 corresponds to the lowest relative drag since it is based on $Ca_{cr}$. A complete dataset of the calculated plate lengths and associated flow velocities are provided for $a_{r} = 100$ and $a_{r} = 1000$ in Table 3 and Table 4, respectively, at $Re = 1$, $Ca_{cr}$ = 27. Overall, the datasets presented here may help to design the experiments to verify the present model.

\subsection{Limitations of the model}
The present model does not account for wake structures past the plate which may form due to vortex shedding at the tip of the plate \cite{Joshi, Soti} and associated flow-induced vibrations (or flutter) of the plate \cite{Schouveiler}. However, the vortex shedding or flutter are absent at low Reynolds number at which the skin friction drag plays a crucial role in determining the drag reduction during flow-induced reconfiguration. For instance, flow-visualization around an elastic fiber reported by Wexler et al. \cite{Wexler} does not show vortex shedding at the tip of the fiber at $Re$ $\sim$ O$(10^{-3})$. The present model does not consider waves in the background flow \cite{Luhar2} as well as the interaction of such pulsatile flows with the thin plate \cite{Kundu}. Three-dimensional effects such as torsional deformation are also not captured by the two-dimensional model presented here \cite{Gosselin2}.

The extension of the plate due to the skin friction drag is ignored in the model and this assumption is valid for thin plates i.e. plates with larger aspect ratio ($a_r$). In order to quantify it, we estimate the maximum extensional strain ($\epsilon$) due to the skin friction drag (shear stress), that is exhibited on the top and bottom surface of the horizontal plate. The shear force per unit plate length ($f_{Ds}$) is obtained using $\theta = 90^o$ in eq. 2. The extensional strain ($\epsilon$) is therefore expressed as follows,
\begin{dmath}
\epsilon = {\dfrac{2f_{Ds}l}{2blG}}{=}\dfrac{\rho C_{Ds} U^{2}}{2G}
\end{dmath}
where $G$ is shear modulus. Simplifying eq. 26 using eqs. 8 and 9, we obtain $\epsilon$ in terms of $Ca$ as follows,
\begin{dmath}
\epsilon = \dfrac{Ca}{a_r^3} \dfrac{(1+\nu)\lambda}{6}
\end{dmath}
where $Ca$, $\nu$, $\lambda$ and $a_r$ are Cauchy number, Poisson ratio, ratio of drag coefficients (eq. 9) and aspect ratio of the plate, respectively. As discussed in section 3.2.2, the plate becomes almost flat at $Ca \approx 10^3$ for $Re = 1$ ($\lambda$ = 0.70). Using these values in eq. 27, the \textcolor{red}{extensional} strain calculated for an incompressible plate ($\nu$ = 0.49) for $a_r = 10$, $a_r = 100$ and $a_r = 1000$ are around $0.17$, $1.7 \times 10^{-4}$ and $1.7 \times 10^{-7}$, respectively. At $a_r \geq 21$, the \textcolor{red}{extensional} strain is estimated as, $\epsilon \leq 0.02$ and therefore, the extension due to the shear force is lesser than 2\% for a plate with aspect ratio larger than 21 and can be neglected in the model.

\section{Conclusions}
The present study investigates the drag reduction on a flexible, thin plate due to the flow-induced reconfiguration in presence of the skin friction drag. The plate is subjected to a free stream uniform flow and is tethered at one end. We extend a model based on Euler-Bernoulli beam theory, reported by Luhar and Nepf \cite{Luhar1}, to account for the skin friction drag on the plate.  In the present study, the range of the Reynolds number ({\it{Re}}) and Cauchy number ({\it{Ca}}) are [${10^{-2}}$, ${10^8}$] and [${10^{-3}}$, ${10^5}$], respectively. 

While the skin friction drag does not play any role for a limiting case of a plate with very large stiffness, \(Ca\) $\rightarrow$ $0$, its influence on a plate with finite stiffness is a strong function of {\it{Ca}} and {\it{Re}}. The skin friction drag is found to be important at ${\it{Ca}}$  ${\geq}$  1 and 0.01 ${\leq}$ $Re$ ${\leq}$ 1. In this range, the total drag on the plate with respect to a rigid upright plate (or total relative drag) decreases due to the flow-induced reconfiguration and further reconfiguration increases the total relative drag due to dominating skin friction drag. A critical Cauchy number (\(Ca_{cr}\)) exists at a given Reynolds number for the lowest relative drag, at which the plate reconfigures itself to a unique shape. $Ca_{cr}$ depends upon the ratio of the form drag coefficient for a upright rigid plate and the skin friction drag coefficient for a horizontal rigid plate (${\lambda}$). In presence of buoyancy, \(Ca_{cr}\) required for achieving the total relative minimum drag is larger.

The modifications in the drag force scaling with free stream velocity ($F_{x}$ ${\propto}$ $U^{n}$) in the presence of the skin friction drag are proposed for Reynolds number, 0.01 $\leq$ \(Re\) ${\leq}$ 1. The present results show that $F_{x}$ ${\propto}$ $U^{4/5+{\lambda}/5}$ for \(1 \leq Ca < Ca_{cr}\) and $F_{x}$ ${\propto}$ $U^{1+{\lambda}/5}$ for \(Ca_{cr} \leq Ca < 300\). Finally, an assessment of the feasibility of experiments is presented and datasets of possible physical systems are provided using the present model. The results presented here may help to design low-drag structures for technical applications as well as may provide insights of locomotion of biological entities at the microscale. 

\section{Acknowledgments}
R.B. gratefully acknowledges financial support by an internal grant from the Industrial Research and Consultancy Centre (IRCC), IIT Bombay. We thank Mr. Vivek Mishra for preliminary contributions to this work and two anonymous reviewers for useful comments. R.B. thanks Profs. Amit Agrawal and Salil Kulkarni at IIT Bombay for useful discussions.

\pagebreak

\section{Tables}
\begin{table*}[h!]
\caption{Nomenclature used in the present paper}
\centering
\begin{tabular}{|p{0.1\linewidth}|p{0.5\linewidth}|}

\hline
 & \\
Symbols & Definition \\
\hline
 & \\
$a_r$ & Aspect ratio of the plate (= $l/t$)\\
$b$ & Width of the plate\\
$B$ & Buoyancy number (eq. 7)\\
$Ca$ & Cauchy Number (eq. 8)\\
$C_{Dp}$ & Form drag coefficient\\
$C_{Ds}$ & Skin friction drag coefficient\\
$E$ & Young's Modulus\\
$f_{Dp}$ & Local form drag force on the plate (eq. 1) \\
$f_{Ds}$ & Local skin friction drag force on the plate (eq. 2) \\
$h$ & Projected height of plate after reconfiguration\\
$h/l$& Area reduction factor after reconfiguration (eq. 16)\\
$I$ & Second moment of inertia of the plate ($=bt^3/12$)\\
$l$ & Length of the plate\\
$l_{e}/l$& Total effective length (eq. 14)\\
$l_{e,p}/l$& Effective length for form drag (eq. 12)\\
$l_{e,s}/l$& Effective length for skin friction factor (eq. 13)\\
$l_{e,str}/l$& Streamlining factor (eq. 15)\\
$Re$ & Reynolds Number (eq. 10)\\
$s$ & Curvilinear coordinate along the plate\\
$t$ & Thickness of the plate\\
$U$ & Free stream velocity\\
$V$ & Restoring force normal to the plate due to stiffness (eq. 4)\\
$\lambda$ & Ratio of $C_{Dp}$ and $C_{Ds}$ (eq. 9)\\
$\theta$ & Angle between plate and vertical at any point\\
$\rho$ & Density of the fluid\\
$\rho_b$ & Density of the plate\\
$\mu$ & Dynamic viscosity\\
& \\
\hline
\end{tabular}

\pagebreak

\end{table*}

\pagebreak

\begin{table*}[h!]
\begin{minipage}{\textwidth}
\centering
\caption{Drag coefficients at different Reynolds numbers obtained from the literature \cite{Vogel_1996,Luhar1,Tom,White}.}
\begin{tabular}{|p{0.14\linewidth}|p{0.15\linewidth}|p{0.15\linewidth}|p{0.15\linewidth}|}
\hline
 & & & \\
Reynolds number \((Re)\) & Form drag coefficient for a upright rigid plate \((C_{Dp})\) & Skin friction drag coefficient for a horizontal rigid plate \((C_{Ds})\) & \(\lambda=C_{Ds}/C_{Dp}\) \\
\hline
 & & & \\
 0.01 & 184.82 & 161.12 & 0.87\\
0.1 & 27.94 & 22.86 & 0.82\\
1 & 5.58 & 3.87 & 0.70\\
10 & 3.80 & 0.65 & 0.17\\
100 & 2.44 & 0.16 & 0.07\\
1000 & 1.95 & 0.04 & 0.02\\
$10^{8}$ & 1.95 & $1.33 \times 10^{-4}$ & {$\approx$}0\\
\hline
\end{tabular}
\end{minipage}
\end{table*}

\begin{table*}[h!]
\begin{minipage}{\textwidth}
\centering

\caption{Plate length (m) and flow velocity (m s$^{-1}$)  for plates of different materials with two fluids at aspect ratio $a_{r}=100$ and $Re = 1$ and $Ca_{cr}$ = 27.} 
\begin{tabular}{|p{0.25\linewidth}|p{0.30\linewidth}|p{0.30\linewidth}|}
\hline
 & &  \\
Plate material $\downarrow$, Fluid$\rightarrow$& Water & Glycerol \\
\hline
 & &  \\
Silicone foam & $5.0\times10^{-5}$ m, $2.0 \times 10^{-2}$ m s$^{-1}$& $6.2\times10^{-2}$ m, $1.8 \times 10^{-2}$ m s$^{-1}$\\
Polyurethane & $6.8\times10^{-6}$ m, $1.5 \times 10^{-1}$ m s$^{-1}$ &  $8.5\times10^{-3}$ m, $1.3 \times 10^{1}$ m s$^{-1}$ \\
HDPE &  $1.2\times10^{-6}$ m, $8.7 \times 10^{-1}$ m s$^{-1}$&  $1.4\times10^{-3}$ m, $7.7 \times 10^{1}$ m s$^{-1}$ \\
Steel &  $7.5\times10^{-8}$ m, $1.3 \times 10^{1}$ m s$^{-1}$ &  $9.4\times10^{-5}$ m, $1.2 \times 10^{0}$ m s$^{-1}$\\
\hline
\end{tabular}

\end{minipage}
\end{table*}

\begin{table*}[h!]
\begin{minipage}{\textwidth}
\centering

\caption{Plate length (m) and flow velocity (m s$^{-1}$)  for plates of different materials with two fluids at aspect ratio $a_{r}=1000$, $Re = 1$ and $Ca_{cr}$ = 27.}
\begin{tabular}{|p{0.26\linewidth}|p{0.30\linewidth}|p{0.30\linewidth}|}
\hline
 & &  \\
Plate material $\downarrow$, Fluid$\rightarrow$& Water & Glycerol \\
\hline
 & &  \\
 Silicone foam & $1.6\times10^{-3}$ m, $6.4 \times 10^{-4}$ m s$^{-1}$ & $2.0\times10^{0}$ m, $5.7\times 10^{-4}$ m s$^{-1}$\\
Polyurethane & $2.2\times10^{-4}$ m, $4.0 \times 10^{-3}$ m s$^{-1}$ &  $2.7\times10^{-1}$ m, $4.1 \times 10^{-3}$ m s$^{-1}$ \\
HDPE &  $3.7\times10^{-5}$ m, $2.7 \times 10^{-2}$ m s$^{-1}$ &  $4.6 \times10^{-2}$ m, $2.4 \times 10^{-2}$ m s$^{-1}$ \\
Steel &  $2.4\times10^{-6}$ m, $4.2 \times 10^{-1}$ m s$^{-1}$  &  $3.0\times10^{-3}$ m, $3.8 \times 10^{-1}$ m s$^{-1}$\\
\hline
\end{tabular}

\end{minipage}

\pagebreak

\end{table*}

\pagebreak

\section{Figures}
\begin{center}
\centering
\includegraphics[width=0.9\columnwidth]{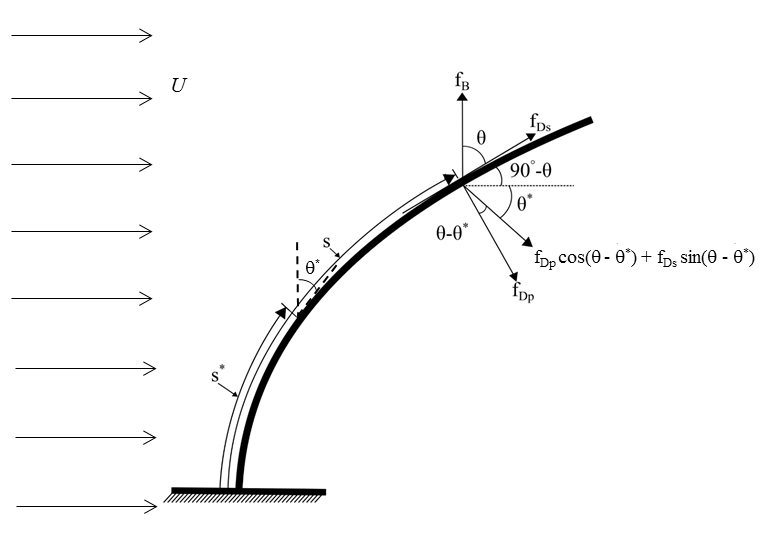}\\
\captionof{figure}{Coordinate system and force components for a reconfigured plate. The contribution of the present work is the inclusion of force due to skin friction drag, $f_{Ds}$, in the model presented by Luhar and Nepf \cite{Luhar1}.}
\end{center}

\pagebreak
\begin{center}
    \centering
 \includegraphics[width=0.7\columnwidth]{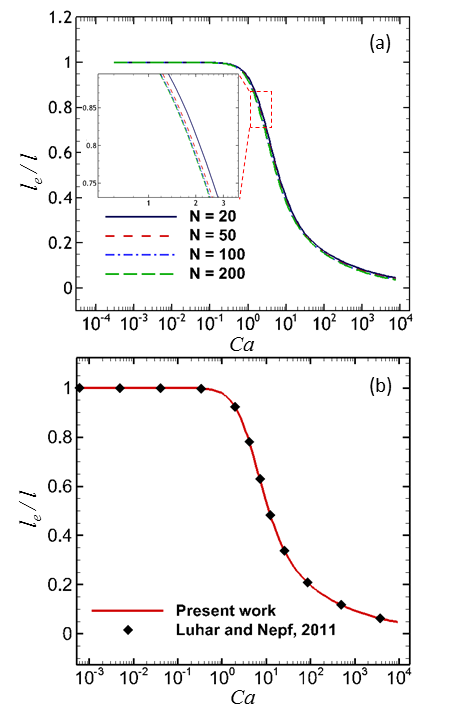}\\
 \captionof{figure}{(a) Grid size convergence test. Effective length (\(l_{e}/l\)) is plotted as function of Cauchy number (\(Ca\)) for different grid points $N$. Inset shows that the results converge for $N \geq 50$. (b) Comparison between the effective length (\(l_{e}/l\)) obtained by the present model and that reported by Luhar and Nepf \cite{Luhar1}. The results in (a) and (b) are obtained at large Reynolds number ($Re = 10^8$ or $\lambda = 0$) and zero buoyancy ($B = 0$).}  
\end{center}

\pagebreak
\begin{center}
    \centering
 \includegraphics[width=0.8\columnwidth]{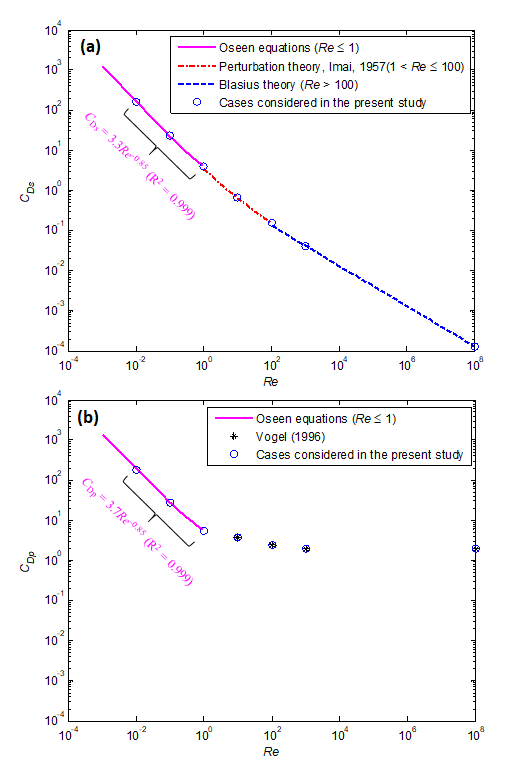}\\
 \captionof{figure}{Variation of skin friction drag coefficient, $C_{Ds}$ (a) and form drag coefficient, $C_{Ds}$ (b), with Reynolds number, $Re$, on a log-log plot. Circles show different cases of $Re$ considered in the present study.}  
\end{center}

\pagebreak
\begin{center}
 \centering
 \includegraphics[width=0.7\columnwidth]{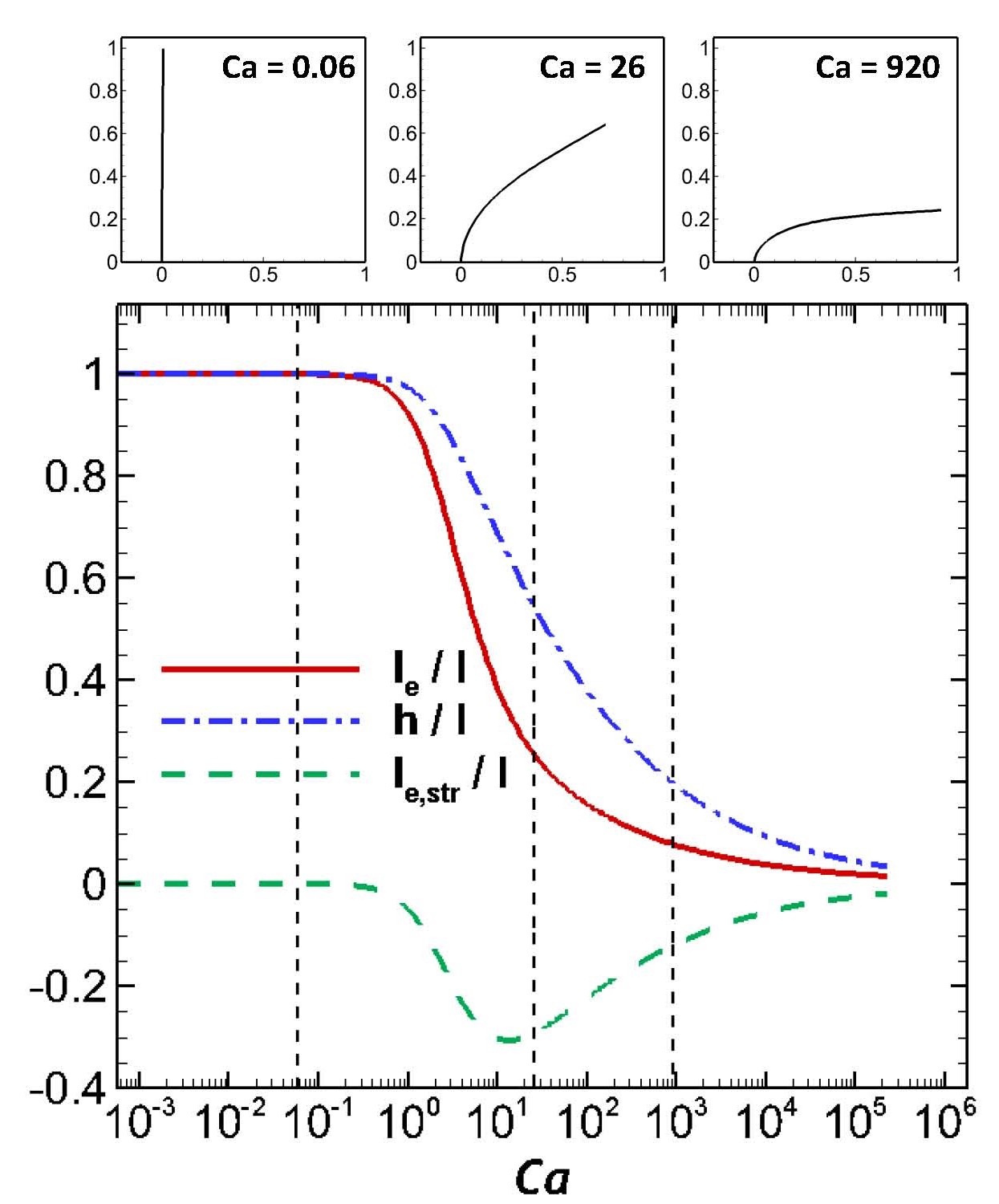}
\end{center}
\captionof{figure}{Area reduction (\(h/l\)), streamlining (\(l_{e,str}/l\)) and total effective length (\(l_{e}/l\)) are plotted as function of Cauchy number (\(Ca\)) in absence of skin friction drag ($\lambda = 0$). The top insets show the shapes of deformed plate at different values of \(Ca\). The values of $Ca$ in the insets are represented by dashed vertical lines in the figure.}

\pagebreak
\begin{center}
 \centering
 \includegraphics[width=0.64\columnwidth]{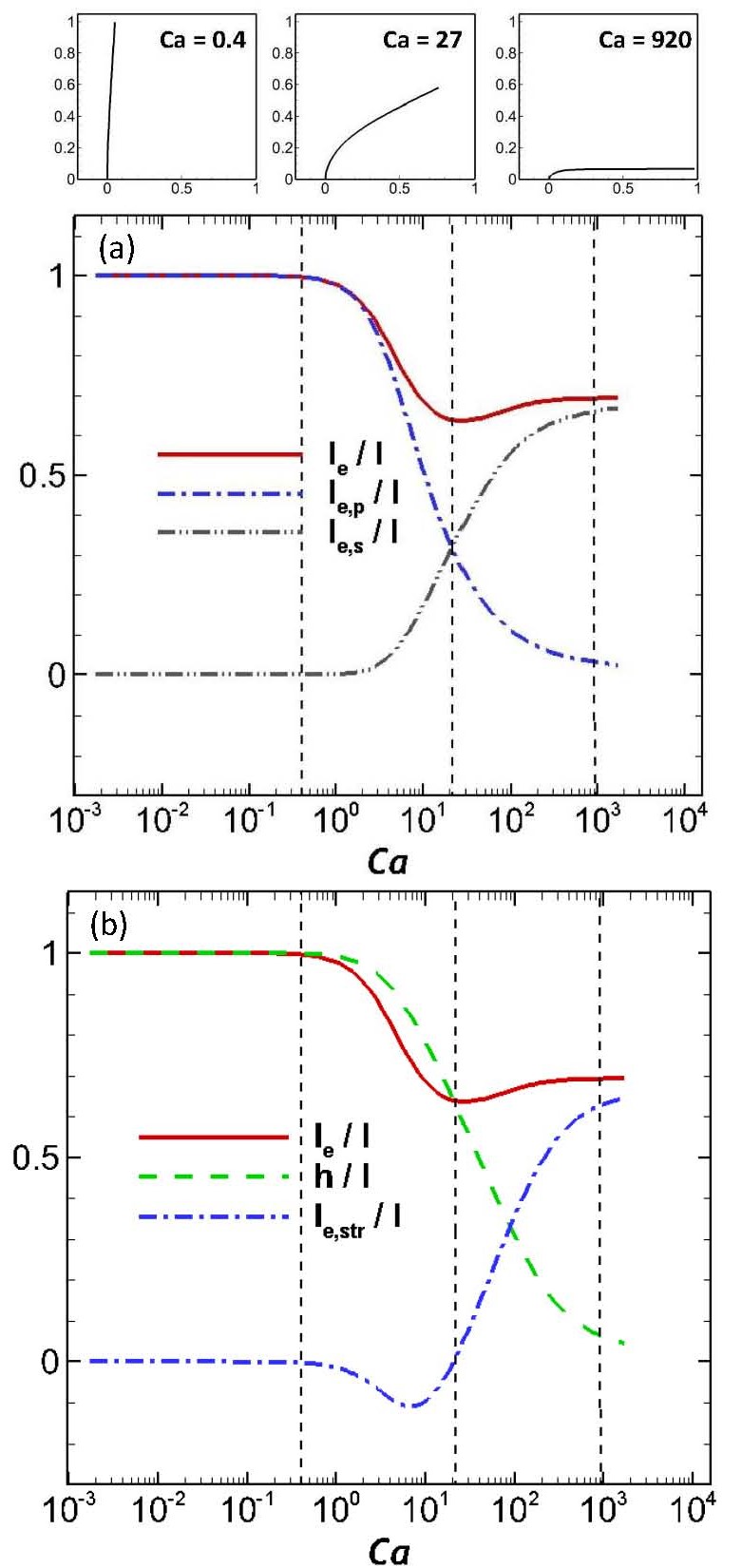}

\end{center}
 \captionof{figure}{(a) Effective length for form drag (\(l_{e,p}/l\)), effective length for skin friction drag (\(l_{e,s}/l\)) and total effective length (\(l_{e}/l\)) are plotted as function of Cauchy number \(Ca\) at Reynolds number, $Re$ = 1. (b) Area reduction (\(h/l\)), streamlining (\(l_{e,str}/l\)) and total effective length (\(l_{e}/l\)) are plotted as function of \(Ca\) at $Re$ = 1. The top insets show the deformed shapes of the reconfigured plate at different values of Cauchy number $Ca$. The values of $Ca$ in the insets are represented by dashed vertical lines in the figures.}

\pagebreak
\begin{center}
 \centering
 \includegraphics[width=0.8\columnwidth]{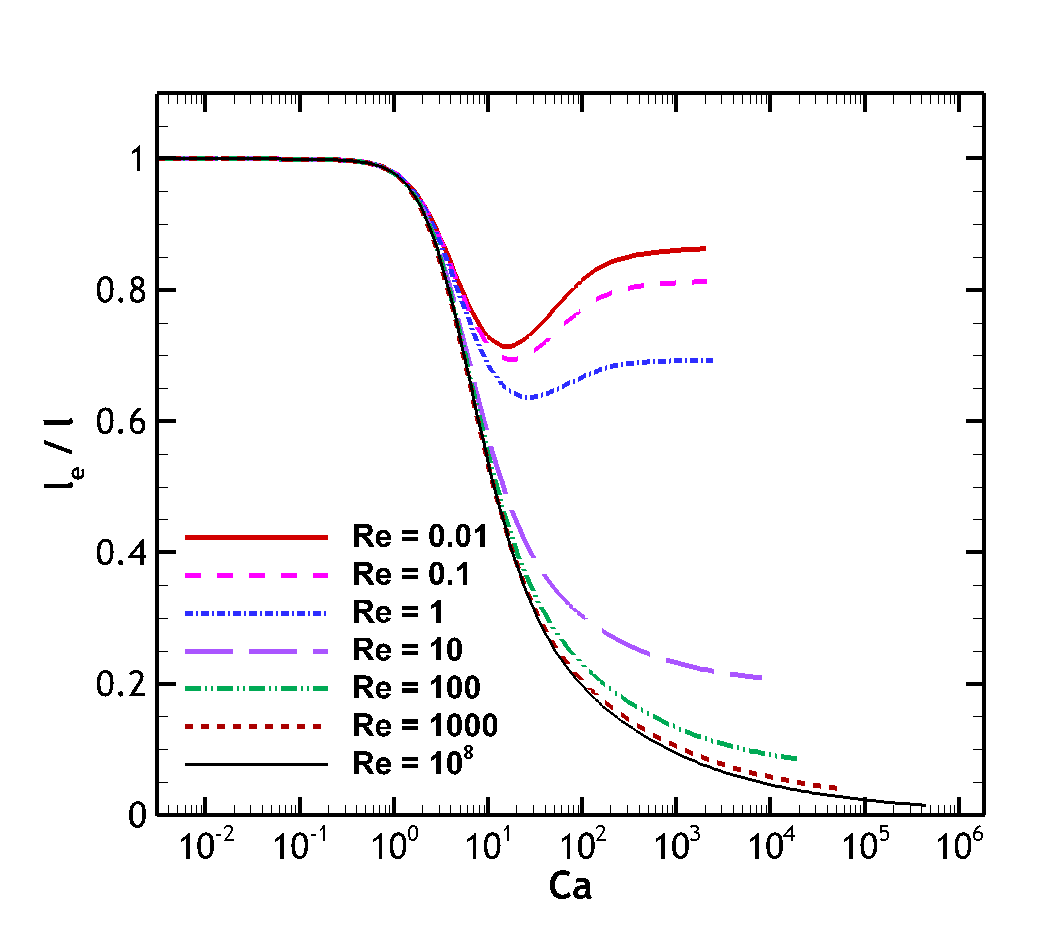}
\end{center}
\captionof{figure}{Variation of total effective length (\(l_{e}/l\)) with Cauchy number (\(Ca\)) for several cases of Reynolds numbers ({\it{Re}}). The variation shows non-monotonic behavior for 0.01 $\leq$ $Re$ $\leq$ 1 as compared to that for 10 $\leq$ $Re$ $\leq$ $10^8$, exhibiting strong influence of skin friction drag in the former.}

\pagebreak
 \begin{center}
 \centering
  \includegraphics[width=0.8\columnwidth]{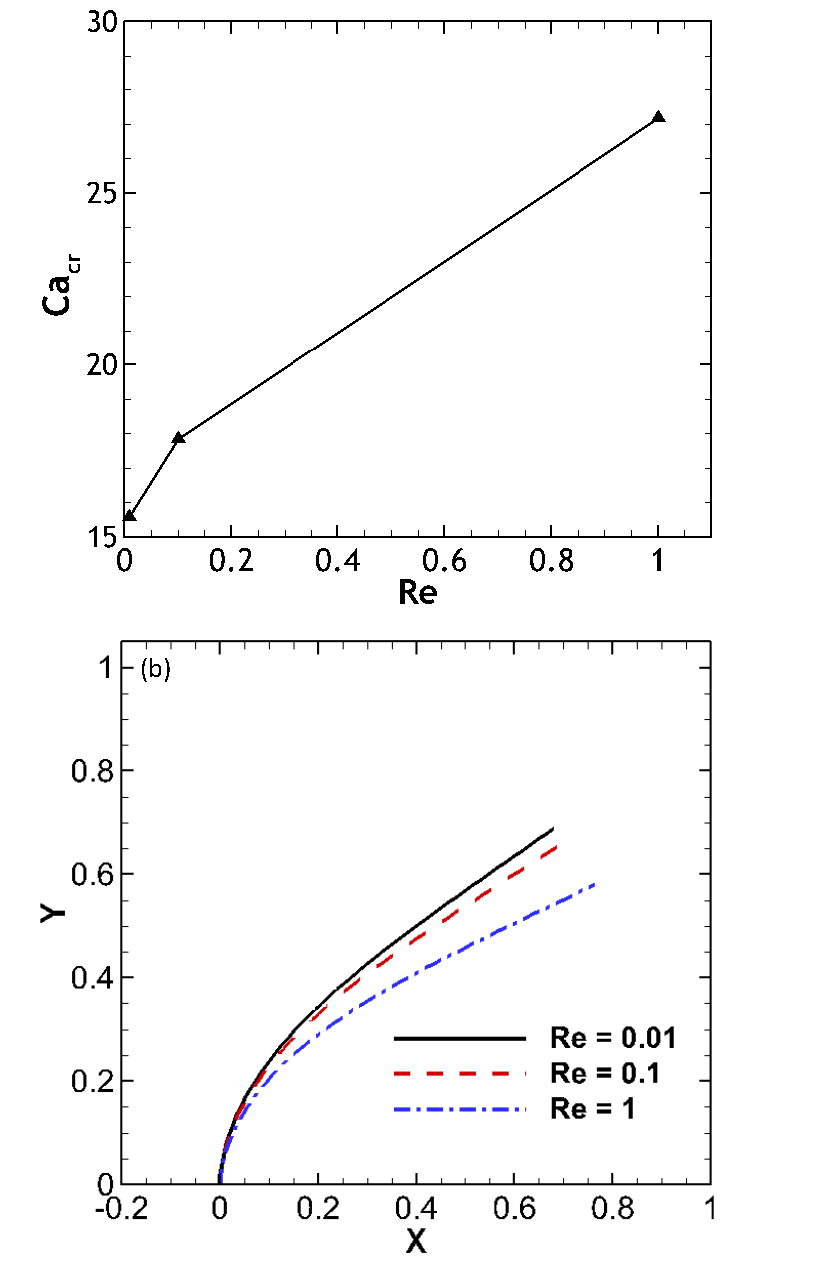}
\end{center}
\captionof{figure}{(a) Critical Cauchy number \(Ca_{cr}\) corresponding to total relative minimal drag are plotted as function of Reynolds number ({\it{Re}}) (a) Shapes of the plate corresponding to the total relative minimal drag are plotted for different Reynolds numbers, {\it{Re}} = 0.01, 0.1 and 1.}

\pagebreak
\begin{center}
    \centering
 \includegraphics[width=0.9\columnwidth]{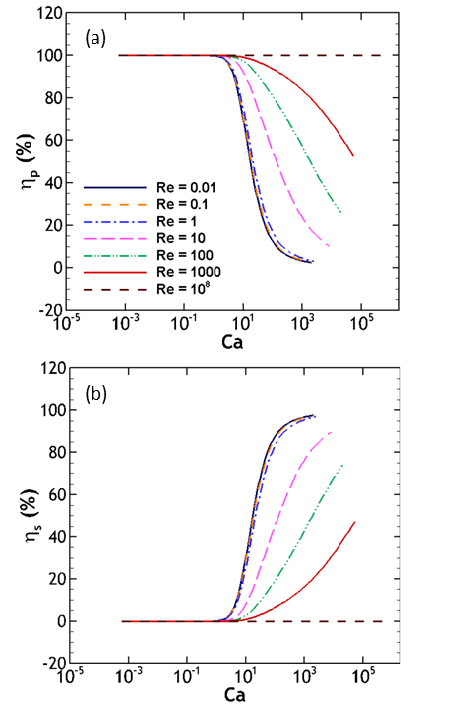}
 \captionof{figure}{Percentage contribution of form drag (a) and skin friction drag (b) in the  total relative drag as function of Cauchy number (\(Ca\)) at different Reynolds numbers ($Re$).}\label{fig1}
\end{center}

\pagebreak
\begin{center}
 \includegraphics[width=0.7\textwidth]{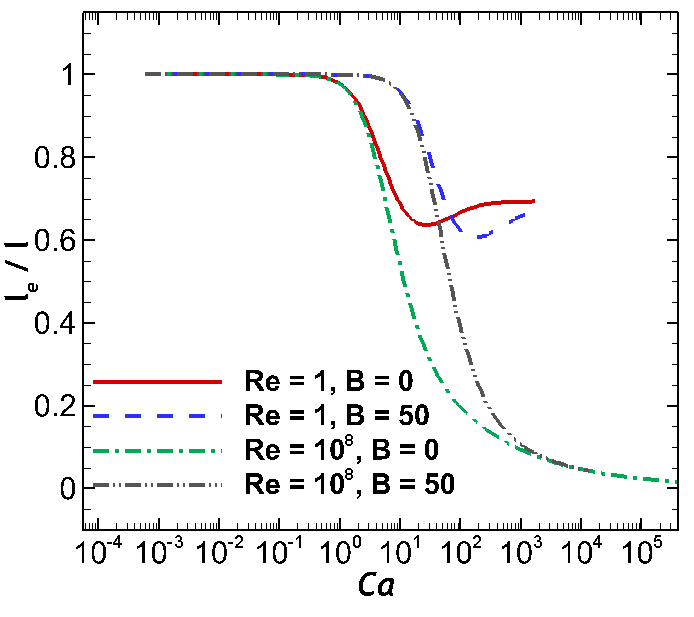}
 \captionof{figure}{Variation of effective length (\(l_{e}/l\)) is plotted as function of Cauchy number (\(Ca\)) for quantifying combined effect of skin friction drag and buoyancy on the drag reduction. Cases with and without buoyancy are considered at lower and larger Reynolds number ($Re$). $B$ is Buoyancy number.}
\end{center}

\pagebreak
\begin{center}
\includegraphics[width=0.8\textwidth]{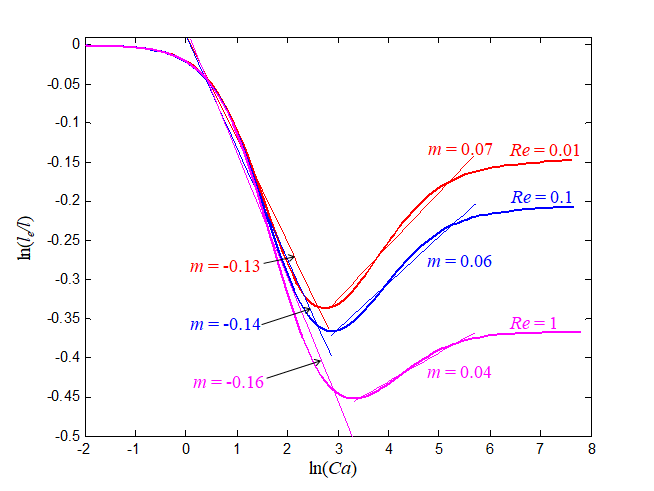}
\captionof{figure}{Power-law fittings for variation of effective length with Cauchy number ($l_{e}/l$ $\propto$ $Ca^m$) are achieved by fitting linear profiles on a log-log plot. The slope of the fit represents the exponent $m$. Fittings are obtained using least-squares method in the two regimes of $Ca$ at different Reynolds numbers, $Re$ = 0.01, 0.1 and 1. $R^2$ values of the plotted fits are larger than 0.98. }
\end{center}

\pagebreak
\begin{center}
\includegraphics[width=0.75\textwidth]{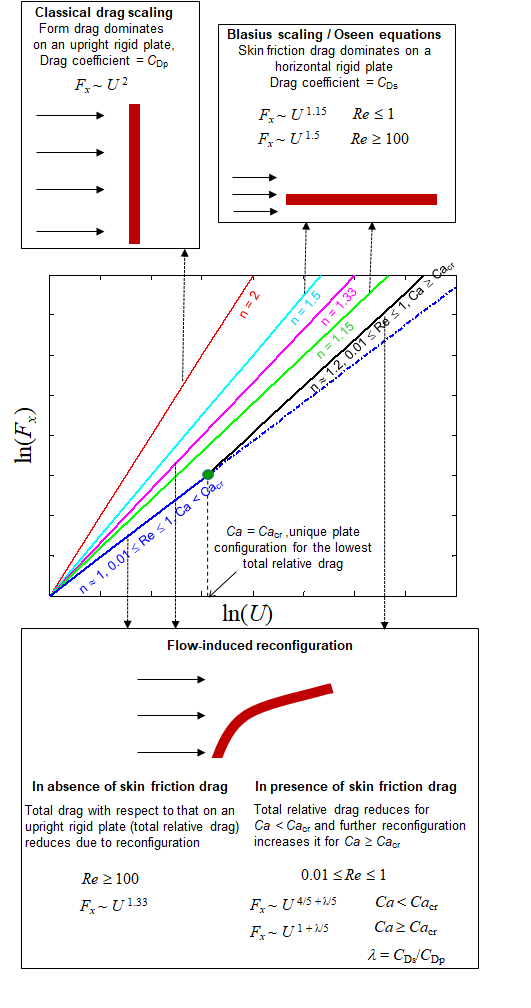}
\captionof{figure}{Comparison among exponent $n$ for scaling of drag force ($F_x$) with free stream velocity ($U$), $F_x$ = $kU^n$, on log-log scale for several possible scenarios. The constant $k$ is same in all cases. The insets show possible plate configuration with the corresponding scaling.}
\end{center}

\pagebreak
\begin{center}
 \includegraphics[width=0.9\textwidth]{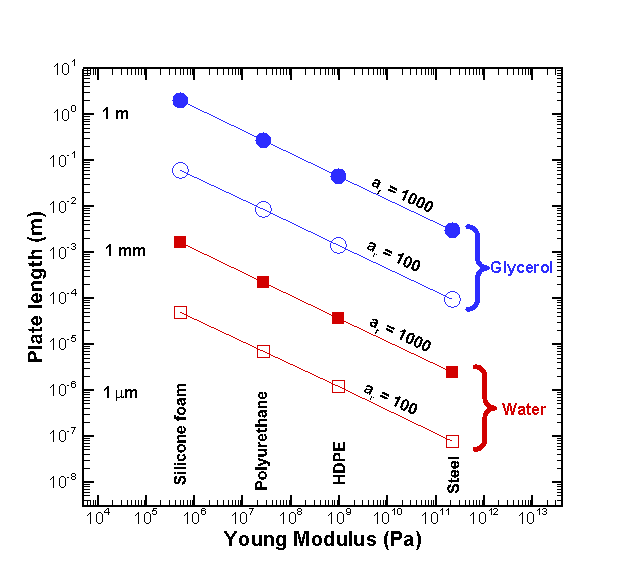}
 \captionof{figure}{Plate lengths are plotted as function of Young's Modulus at critical Cauchy number. Combinations of four plate materials - silicone foam, polyurethane, HDPE and steel - and two fluids - water and glycerol - are considered.  The plates of aspect ratios, $a_{r} = 100$ and $a_{r} = 1000$ are plotted with hollow and filled symbols, respectively.} 
\end{center}

\end{large}


\begin{thebibliography}{99}

\bibitem{Alben1} Alben S., M. Shelly, and J. Zhang, Drag reduction through self-similar bending of flexible body, {\it{Nature}}, vol. 120, p. 3, 2002.

\bibitem{Alben2} Alben S., M. Shelley, J. Zhang, How flexibility induces streamlining in a two-dimensional flow?, {\it{Physics of Fluids}}, Vol. 16, p1694, 2004.  

\bibitem{Langre} de Langre E. , Effects of wind on plants, {\it{Annual Review of Fluid Mechanics}}, Vol. 40, p141, 2008.

\bibitem{Langre2} de Langre E., A Gutierrez, J Cossé, On the scaling of drag reduction by reconfiguration in plants, {\it{Comptes Rendus Mécanique}}, Vol.340, p35, 2012

\bibitem{Gosselin} Gosselin F., E. de Langre, B.A. Machado-Almeida “Drag reduction of flexible plates by reconfiguration” {\it{Journal of Fluid Mechanics}}, Vol. 650, p319, 2010.

\bibitem{Gosselin2} Hassani M., Mureithi, N., Gosselin, F.P., Large Coupled Bending and Torsional Deformation of an Elastic Rod Subjected to Fluid Flow, {\it{Journal of Fluids and Structures}}, Vol 62, p367, 2016.

\bibitem{Hen} Henriquez S., A. Barrero-Gil, Reconfiguration of flexible plates in sheared flow, {\it{Mechanics Research Communications}}, Vol. 62, p1, 2014.

\bibitem{Imai} Imai I., Second Approximation to the Laminar Boundary Layer Flow over a Flat Plate, {\it{Journal of Aeronautical Science}}, Vol. 24, pp 155-156, 1957.

\bibitem{Joshi} Joshi R. U., Soti A, Bhardwaj R, Numerical study of Heat Transfer Enhancement by Deformable Twin Plates in Laminar Heated Channel flow, {\it{Computational Thermal Sciences}}, Vol. 7, pp 1-10, 2015.

\bibitem{Kundu} Kundu A. K., Soti A. K. Bhardwaj R., Thompson M., The Response of an Elastic Splitter Plate Attached to a Cylinder to Laminar Pulsatile Flow, {\it{Journal of Fluids and Structures}}, Vol. 68, Pages 423–443, 2017.

\bibitem{Leclercq} Leclercq T., E. de Langre, Drag reduction by elastic reconfiguration of non-uniform beam in non-uniform flows, {\it{Journal of Fluid and Structures}}, Vol. 60, pp. 114-129, 2016.

\bibitem{Shelley2} Lindner A. and M. Shelley, Elastic fibers in flows in Fluid-structure interactions at low Reynolds numbers, (eds. C. Duprat and H. A. Stone), {\it{Royal Society of Chemistry}}, 2016.

\bibitem{Luhar1} Luhar M., H. M. Nepf, Flow-induced reconfiguration of buoyant and flexible aquatic vegetation, {\it{Limnol. Oceanogr.}}, Vol. 56, p2003, 2011. 

\bibitem{Luhar2} Luhar M., H. M. Nepf, Wave-induced dynamics of flexible blades, {\it{Journal of fluids and structures}}, Vol. 61, p20, 2016.

\bibitem{Purcell} Purcell E. M., Life at low Reynolds number, {\it{American Journal of Physics}} Vol. 45, p3, 1977.

\bibitem{Schouveiler} Schouveiler L., C. Eloy, and P. Le Gal, Flow-induced vibration of high mass ratio flexible filaments freely hanging in a flow, {\it{Physics of Fluids}}, Vol. 17, p. 047104, 2005.

\bibitem{Shelley1} Shelley M., J. Zhang, Flapping and Bending Bodies Interacting with Fluid Flows, {\it{Annual Review of Fluid Mechanics}} Vol. 43, 449-465, 2011.

\bibitem{Soti} Soti A. K., Bhardwaj R., Sheridan J., Flow-induced Deformation of a Elastic Plate as Manifestation of Heat Transfer Enhancement in Laminar Channel Flow, {\it{International Journal of Heat and Mass Transfer}}, Vol. 84, pp 1070-1081, 2015

\bibitem{Tom} S. Tomotika, T. Aoi, The Steady Flow of a Viscous Fluid Past an Elliptic Cylinder and a Flat Plate at Small Reynolds Numbers, {\it{Quarterly Journal of Mechanics and Applied Mathematics}}, Vol. 6, p3, 1953.

\bibitem{Vogel_1996} Vogel, S., Life in Moving Fluids: The Physical Biology of Flow, {\it{Princeton Universsity Press}}, 1996. Pg 98. 

\bibitem{Zhu} Zhu L., Scaling laws for drag of a complaint body in an incompressible viscous flow, {\it{Journal of fluid mechanics}}, Vol. 607, p387, 2008. 

\bibitem{White} White F. M., Viscous Fluid Flow, {\it{McGraw Hill, New York}}, 3rd Edition, 2005. 

\bibitem{Wexler} Wexler J. S., P. H. Trinh, H. Berthet, N. Quennouz, O. d. Roure, H. E. Huppert, A. Linder, H. A. Stone, Bending of elastic fibres in viscous flows: the influence of confinement, {\it{Journal of Fluid Mechanics}}, Vol. 733, p684, 2013.


\end{thebibliography}
\end{document}